\documentclass[preprint2]{aastex}

\usepackage{color}

\let\underscore\_
\newcommand{\myunderscore}{\renewcommand{\_}{\underscore\hspace{0pt}}}
\myunderscore

\slugcomment{Submitted to AJ, \today}

\shorttitle{Tidal Dwarf Galaxies of NGC4922}
\shortauthors{Sheen et al.}

\begin{document}


\title{TIDAL DWARF GALAXIES AROUND A POST-MERGER GALAXY, NGC\,4922}


\author{Yun-Kyeong Sheen\altaffilmark{1}, Hyunjin Jeong\altaffilmark{1},
Sukyoung K. Yi\altaffilmark{1}$^{\star}$, Ignacio Ferreras\altaffilmark{2}, 
Jennifer M. Lotz\altaffilmark{3}, Knut A. G. Olsen\altaffilmark{3}, Mark Dickinson\altaffilmark{3},
Sydney Barnes\altaffilmark{4}, Jang-Hyun Park\altaffilmark{5}, Chang H. Ree\altaffilmark{5},
Barry F. Madore\altaffilmark{6}, Tom A. Barlow\altaffilmark{7}, Tim Conrow\altaffilmark{7}, Karl Foster\altaffilmark{7}, 
Peter G. Friendman\altaffilmark{7}, Young-Wook Lee\altaffilmark{1}, D. Christopher Martin\altaffilmark{7}, 
Patrick Morrissey\altaffilmark{7}, Susan G. Neff\altaffilmark{8}, 
David Schiminovich\altaffilmark{9}, Mark Seibert\altaffilmark{6}, 
Todd Small\altaffilmark{7}, Ted K. Wyder\altaffilmark{7}}
\email{yi@yonsei.ac.kr}


\altaffiltext{1}{Department of Astronomy, Yonsei University, Seoul
  120-749, Republic of Korea}
\altaffiltext{2}{Mullard Space Science Laboratory, University College London,
Holmbury St Mary, Dorking, Surrey, RH5 6NT, UK}
\altaffiltext{3}{National Optical Astronomy Observatory, 950 North
  Cherry Avenue, Tucson, AZ 85719, USA}
\altaffiltext{4}{Lowell Observatory, Flagstaff, AZ 86001, USA}
\altaffiltext{5}{Korea Astronomy and Space Science Institute, Hwaam
  61-1, Yuseong, Daejeon 305-348, Republic of Korea}
\altaffiltext{6}{Observatories of the Carnegie Institution of
  Washington, 813 Santa Barbara St., Pasadena, CA 91101, USA}
\altaffiltext{7}{California Institute of Technology, MC 405-47, 1200
  E. California Blvd., Pasadena, CA 91125, USA}
\altaffiltext{8}{Laboratory for Astronomy and Solar Physics, NASA
  Goddard Space Flight Center, Greenbelt, MD 20771, USA}
\altaffiltext{9}{Department of Astronomy, Columbia University, New
  York, NY 10027, USA}


\begin{abstract}
One possible channel for the formation of dwarf
galaxies involves birth in the tidal tails of interacting galaxies. 
We report the detection of a bright UV tidal tail and several
young tidal dwarf galaxy (TDG) candidates
in the post-merger galaxy NGC\,4922 in the Coma cluster.
Based on a two-component population model (combining young and 
old stellar populations), we find that the light of tidal tail predominantly
comes from young stars (a few Myr old). The {\it Galaxy Evolution
Explorer} ({\it GALEX}) ultraviolet data played a critical role in
the parameter (age and mass) estimation. Our stellar mass estimates
of the TDG candidates are $\sim10^{6-7} M_{\sun}$,
typical for dwarf galaxies.
\end{abstract}





\section{INTRODUCTION}

A tidal dwarf galaxy (hereafter TDG) is defined as ``a dwarf-sized
self-gravitating object assembled from tidal debris''
\citep{hib05}. 
The formation of dwarf galaxies from tidal interactions
was already suggested by \citet{zwi56}.
Two decades later, blue sources were found in the tidal tail of NGC\, 4038/39 
(``the Antennae''), as well as H{\scriptsize I} tails in several
other interacting galaxies \citep{sch78}.
In the 1990s, more TDGs were discovered
\citep{sch90,mir91,mir92}.  During the last two decades, about 20 more
interacting galaxies and galaxy groups were investigated for the
formation of TDGs
\citep{hib94,yos94,duc97,duc98,duc00,hei00,wei00,hib01,men01,wei03,
tem03,mun04,hib05,nef05,boq07,boq09,han09,kor09,smi09,kon09}. TDGs were found to feature
blue optical colors revealing the existence of young stellar
populations. They are found at the tip of tidal tails, which
usually correlates with H{\scriptsize I} gas density peaks. They also show higher
metallicity than typical dwarf spheroidal galaxies, as expected, since 
they are assembled from recycled materials ejected by their merging
parents.

The formation of tidal tails during the encounter of disk galaxies has
been reproduced by numerical simulations \citep[e.g.,][]{too72,bar88,wet07}.
\citet{wal90} explored the triggering of star formation in tidal
tails by combining a dynamical model with a code tracking the color
evolution. Soon after, more detailed numerical simulations of
the formation of dwarf galaxies in tidal tails followed
\citep{bar92,elm93}. More recent simulations achieve much higher resolutions
and include chemical evolution \citep[e.g.,][]{bou08,rec07}.
For instance, \citet{bou08} achieve resolutions high enough to probe down to 
the masses of star clusters ($\gtrsim 10^5 M_\sun$).

Nevertheless, much debate still exists on the fraction of dwarf
galaxies born from tidal interactions, because it is difficult
to identify TDGs once the tidal tail fades away. We can only
identify TDGs by the presence of tidal tails or by a
spectroscopic survey around galaxy mergers. Within the standard
hierarchical formation scenario, fewer young tidal dwarf galaxies
should be expected at present, compared to higher redshifts.
Consequently, the number of confirmed TDGs is too small to
characterize the population or to provide information for numerical
studies.

Here, we focus on the formation of TDGs in the tidal tail of
NGC\,4922.  NGC\,4922 (M$_{\rm r} = -22.1$, z$\sim0.0235$) is a
post-merger galaxy located at the outskirts of the Coma cluster. It is
known as a merger between an early-type galaxy and a spiral
galaxy. Two nuclei from its progenitors can be resolved,
separated by $\sim$10~kpc. This galaxy is especially interesting
because the northern nucleus is a weak active galactic nucleus (AGN), 
probably as a consequence of the
merging event. \citet{and86} and \citet{alo99} identified the northern nucleus as
a Seyfert 2 galaxy with low ionization level. \citet{alo99}
also showed the extended soft X-ray emission which did not originate
from the AGN, and possibly relates to ongoing star formation.

In this paper, we present and discuss a UV-bright tidal tail
of NGC\,4922 and TDG candidates obtained from data taken by the  {\it
Galaxy Evolution Explorer} ({\it GALEX}). 
In Section~\ref{sec:obs}, we describe the {\it GALEX} observations and data
reduction with supporting optical imaging and data reduction. In
Section~\ref{sec:model}, we present the synthetic stellar population
modeling. Star formation histories on the tidal tail and the TDG 
candidates are quantified and discussed in Section~\ref{sec:discussion}. 
We summarize our findings in Section~\ref{sec:summary}.

\begin{figure*}[!ht]
\includegraphics[scale=0.44]{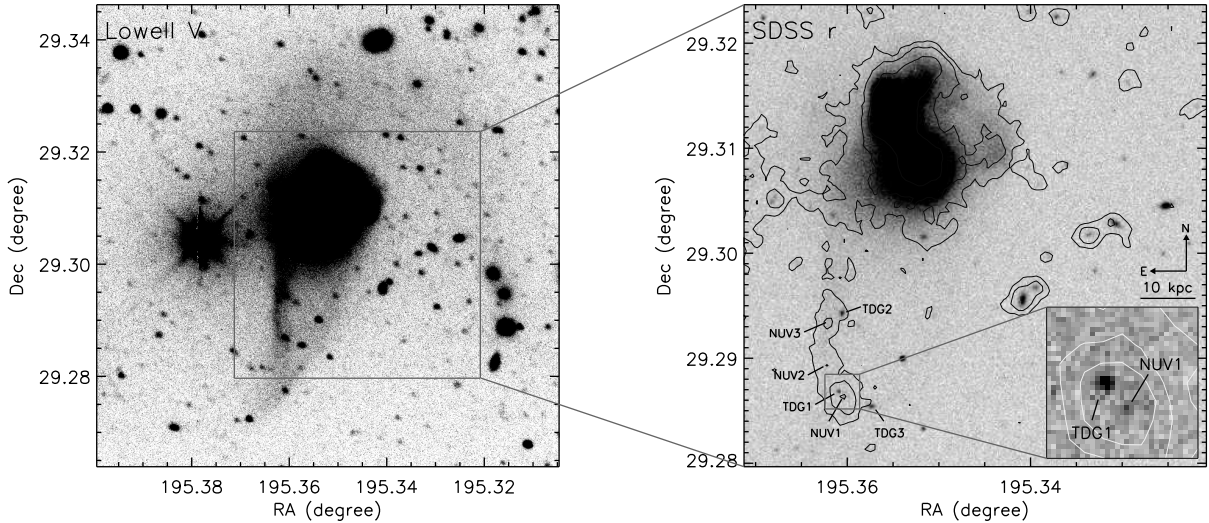}
\caption{Left: optical $V$ band deep image taken at Lowell Observatory. 
In addition to the UV-bright tidal tail, we can see faint tidal structures around 
NGC\,4922. Right: {\it GALEX} NUV surface brightness contours are imposed 
over the optical SDSS r band image. Optical objects and NUV peaks are indicated and
contours are given for $\mu_{\rm NUV}\sim$26.0, 27.0, 28.0~mag/arcsec$^2$. The inset
shows optical counterparts of `TDG1' and `NUV1' in the SDSS r
band. Contours are drawn for $\mu_{\rm NUV}\sim$26.5, 27.0, 28.0~mag/arcsec$^2$ in this
case to see the faint object corresponding to `NUV1'. Because of the poor
resolution of the {\it GALEX} data, we could not deblend the NUV light
into the two optical objects. \label{uvknots}}
\end{figure*}

\section{OBSERVATIONS AND DATA REDUCTION}
\label{sec:obs}

NGC\,4922 is included in the Sloan Digital Sky Survey
\citep[SDSS;][]{yor00}. The typical exposure time for all SDSS {\it u,
  g, r, i,} and {\it z}
bands is 53.9~s and the pixel scale is 0$\farcs$396 pixel$^{-1}$. We
determined FWHM in each band using the \texttt{psfmeasure} task in
Image Reduction and Analysis Facility (IRAF), as
1$\farcs$28 for {\it u}, 1$\farcs$28 for {\it g}, 1$\farcs$09 for {\it
r},
1$\farcs$05 for {\it i} and 0$\farcs$97 for {\it z} band.

In order to explore faint features produced through the tidal
interaction, we carried out optical deep imaging of NGC\,4922 at
Lowell Observatory. Observations were performed on 2007 April 16 with
the Perkins Re-Imaging SysteM (PRISM) mounted on the Perkins 1.8 m
telescope. The field of view of the CCD is
13$\farcm$3 $\times$ 13$\farcm$3 and the pixel scale is 0$\farcs$390
pixel$^{-1}$. Images were taken in the $V$ band and reduced using
IRAF. 
In order to see faint structures around NGC\,4922, one 30
minute exposure and two 20 minute exposures were combined in the $V$
band. We followed the usual procedure for calibration using
standard stars observed in the same night. The FWHM of the combined
image is 2$\farcs$88, also obtained via the {\tt
psfmeasure} task.

We used the {\it GALEX} public data of NGC 4922, taken between 
2005 and 2006 in both the FUV ($1350$--$1750$\AA) and NUV
($1750$--$2750$\AA) bands. The data were acquired through the {\it GALEX} guest
investigation project of Giuseppe Gavazzi \citep[$GI1$\_039006\_Coma\_MOS06;][]
{cor08}.
Details of the {\it GALEX} instruments, 
pipeline and calibration are described in \citet{mar05} and \citet{mor07}. 
Total exposure times were 1693~s and 3524~s in FUV and NUV, respectively.
The spatial resolution of the images are approximately 4$\farcs$2
and 5$\farcs$3 FWHM in FUV and NUV, respectively, sampled with
1$\farcs$5 $\times$ 1$\farcs$5 pixels.

\begin{figure}
\includegraphics[scale=0.45]{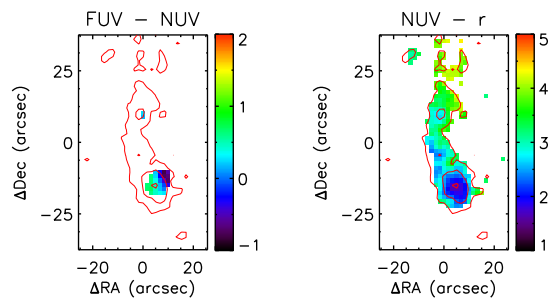}
\caption{Color maps derived from the pixel magnitude maps of the
tidal tail region. S/N$_{\rm FUV} > 1$ cut is applied for `FUV$-$NUV' color
map while S/N$_{\rm NUV} > 1$ cut is used for `NUV $-$ {\it r}' color map.
These maps show blue diagnostic colors
indicating the presence of young stellar populations along the tidal
tail.  \label{colormap}}
\end{figure}

\begin{figure}[!ht]
\includegraphics[scale=0.45]{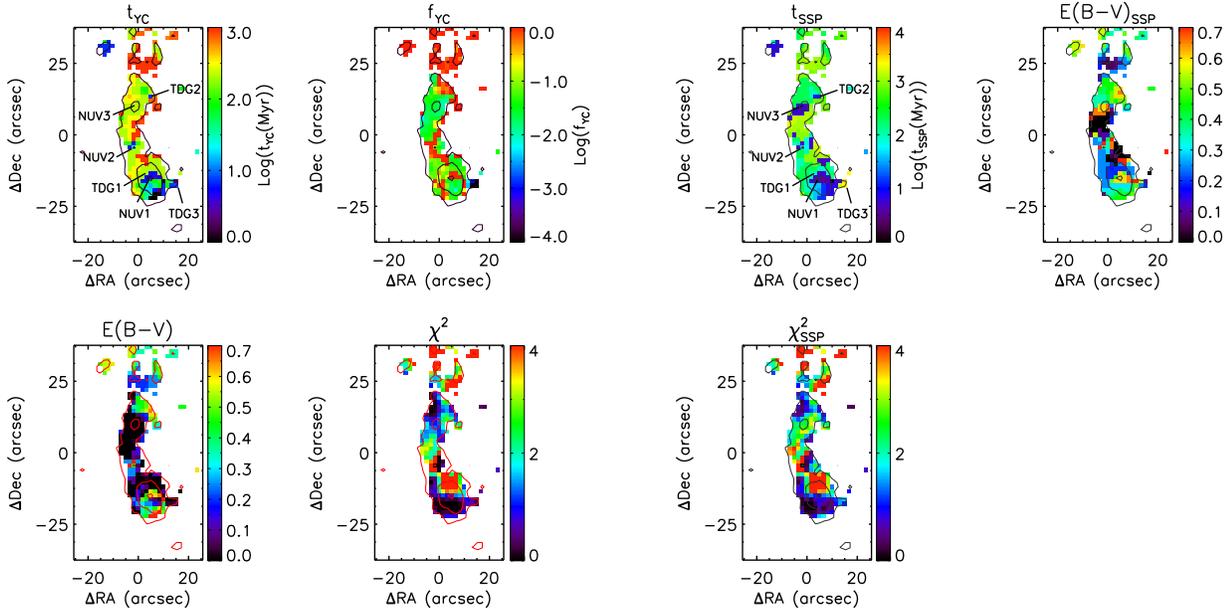}
\caption{Predictions of the pixel-by-pixel analysis with two-component
modeling in the tidal tail region. Only pixels of S/N$_{\rm NUV} > 1$ and 
$\chi^2 < 4$ are presented. NUV surface brightness contours are given for
$\mu_{\rm NUV}=26, 27, 28$~mag/arcsec$^2$. Clockwise from upper left,
each panel shows the age of the young component in units of
Myr, the young stellar mass fraction, $\chi^2$, and the 
internal extinction in $E(B-V)$. 
\label{pixelmap}}
\end{figure}

\begin{figure}[!t]
\includegraphics[scale=0.45]{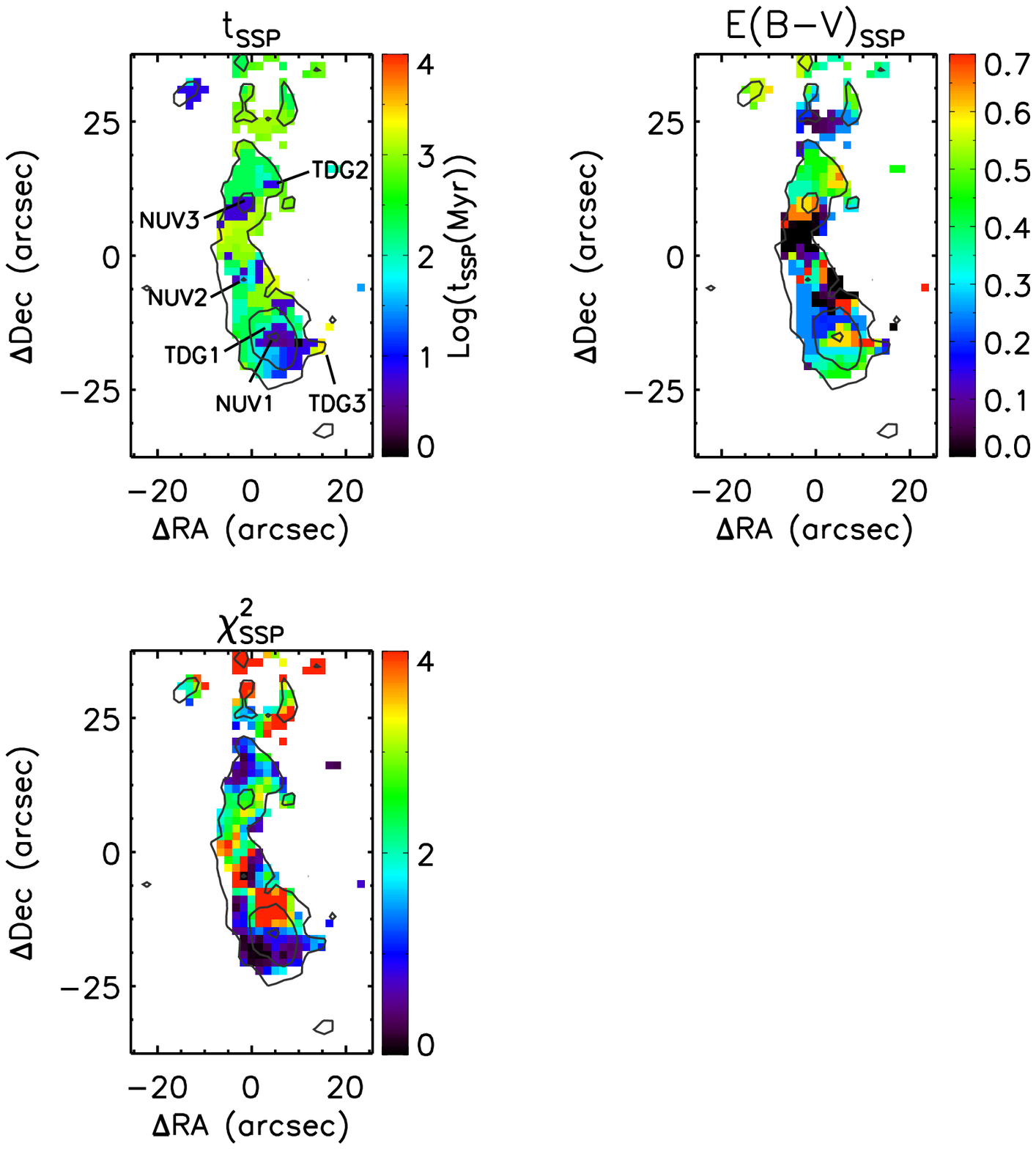}
\caption{Results from SSP modeling of the pixel fluxes in
the tidal tail region. Only pixels of S/N$_{\rm NUV} > 1$ and  $\chi^2 < 4$ are
presented. NUV surface brightness contours are shown at the $\mu_{\rm NUV}=26, 27, 28$~mag/arcsec$^2$.
\label{pixelmap_ssp}}
\end{figure}

Figure~\ref{uvknots} shows the optical images with UV magnitude contours.
The $V$-band deep image in the left panel reveals faint structures around NGC\,4922
to the north and south of the galaxy. Along
the tidal tail, the stellar debris of the interaction are visible.
The right panel shows the shallower SDSS $r$-band image over which the NUV surface 
brightness contours are superimposed for $\mu_{\rm NUV} = 26.0, 27.0, 28.0$~mag/arcsec$^2$.
`TDG1' and `NUV1' are zoomed in in the inset image of the right panel with the NUV surface 
brightness contours for $\mu_{\rm NUV} = 26.5, 27.0, 28.0$~mag/arcsec$^2$.

The apparent magnitudes are taken from the AUTO magnitudes
measured by SExtractor \citep{ber96}. All UV and optical 
magnitudes in this study are measured in AB magnitude system.
Galactic extinction was corrected using the reddening maps of 
\citet*{schetal98} and the extinction law of \citet{car89}. The extinctions for 
the {\it GALEX} passbands are $A_{\rm FUV}=8.24 \times\,E(B-V)$ and $A_{\rm
NUV} = 8.24 \times\,E(B-V)-0.67 \times\,E(B-V)^2$ \citep{ree07}. Our photometric errors
include photon-statistic errors as well as the
flat-field variations of the UV data. In optical data, dark current and
sky background were considered in the error budget.

In order to avoid spurious color gradients in the pixel-by-pixel
analysis, caused by the different point-spread functions, we convolved the UV and optical
data to a common FWHM of 6$\arcsec$ using the circular Gaussian kernel
of {\tt gauss} task in IRAF.

\section{STELLAR POPULATION MODELING}
\label{sec:model}

The main focus of this paper is the study of the underlying
stellar populations in the tidal tail of NGC\,4922, using {\it GALEX} UV and
SDSS optical data. Recent star formation is modeled by a
superposition of an old and a young stellar component, following
previous works \citep[see e.g.][]{fs00,yi05,kav07}. The model takes two simple
stellar populations (SSPs), whereby the free parameters are the age of the
old and young components ($t_{OC}$ and $t_{ YC}$,respectively), the
young stellar mass fraction ($f_{YC}$), and internal extinction
($E(B-V)$).  The models of \citet{yi03} are used to describe the old
populations (1--12 Gyr). The models are based on single stellar
populations with a solar metallicity. Since those
models do not cover ages younger than 1~Gyr, we combine them with the
models of $\it{Starburst99}$ \citep{lei99} at the age of 1~Gyr. The
second burst has a solar metallicity too. Internal extinction is
applied to the composite spectral energy distribution (SED) models. 
The extinction curve is taken from \citet{cal01}.

We construct a library of over half a million SEDs for a grid of
two-component models at the redshift of NGC\,4922 ($z=0.0235$). The
grid has dimension $12 \times 36 \times 38 \times 33$ for $t_{OC}$,
$t_{YC}$, $f_{YC}$ and $E(B-V)$, respectively. $t_{OC}$ extends from
1~Gyr to 12~Gyr while $t_{YC}$ ranges from 1~Myr to 900~Myr. The young
mass fraction, $f_{YC}$, is explored over its full possible range,
0 through 1, and $E(B-V)$ spreads over 0.0 to 0.7.
The grid points in the two-dimensional parameter space spanned by $t_{YC}$ and $f_{YC}$ was
chosen with logarithmic spacings, and adapted to extract the maximum
sensitivity from the photometry on the presence of young populations.

The photometric SEDs of the objects are compared with the SEDs in the
library and the best fit is determined via a standard likelihood
estimator.  We also compare the observations with
models of SSPs, i.e., assuming a single age
and metallicity.

\section{DISCUSSION}
\label{sec:discussion}

\begin{figure}[!ht]
\includegraphics[scale=0.4]{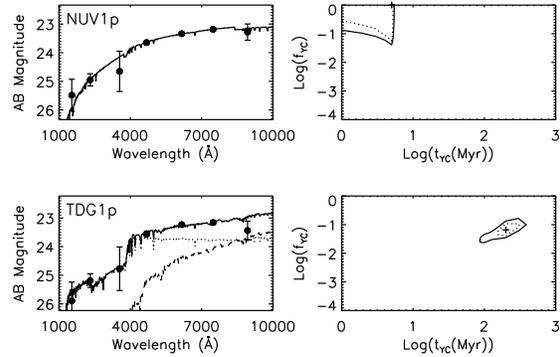}
\caption{Best fit models from the pixel SEDs of ``NUV1p'' and ``TDG1p''
in the left panels. A dotted line shows the model SED of young component,
while the dashed line indicates the contribution of an old component. 
In the right panels, $\chi^2$ contours are presented for the 68\% and
90\% confidence levels (dashed and solid lines, respectively).
A cross indicates the location where a minimum $\chi^2$ was found in
the parameter space. The best-fit values are presented in
Table~\ref{result_tdg}. \label{pixelgraph}}
\end{figure}

\begin{figure}[!t]
\includegraphics[scale=0.45]{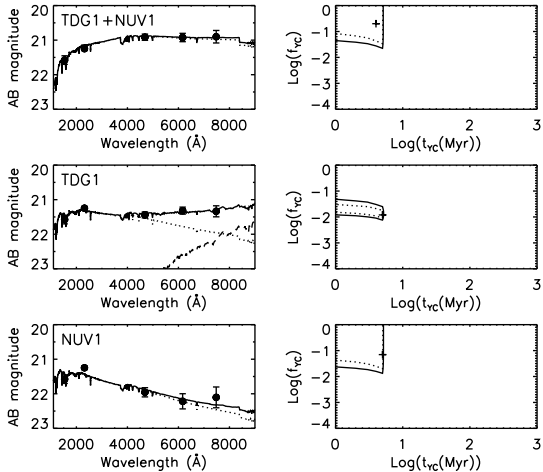}
\caption{Result of stellar population modeling with SEDs derived from
aperture photometry. Since we could not deblend the UV light around ``TDG1''
and ``NUV1'' into two separate sources, we add the optical SEDs
from both objects. The upper panels show results for ``TDG1+NUV,'' the middle
panels are for the optical SED of ``TDG1,'' and the bottom panels are for ``NUV1,'' In
the left panels dotted lines indicate the contribution of a young component while the
dashed line shows a SED of an old component. In the right panels, $\chi^2$ contours
are presented for 68\% and 90\% confidence levels (dashed and solid lines,
respectively). The result shows young populations of $\sim5$~Myr. \label{tdgsed}}
\end{figure}

\begin{figure}[!t]
\includegraphics[scale=0.4]{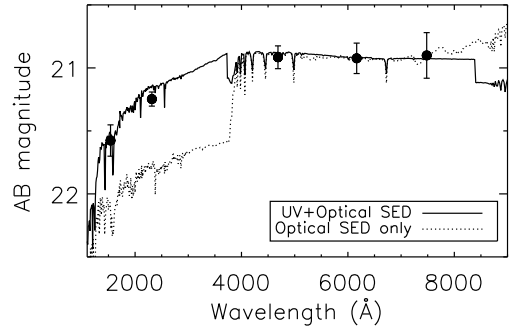}
\caption{Best SSP model fits for ``TDG1+NUV1.'' The best fit is achieved
  for $t_{\rm SSP}=3$~Myr
with the UV-optical SED, while it is 16 Myr when only the optical data are used to constrain the models
(albeit with a large degeneracy). It shows the power of UV data to constrain the
age derivation of young stellar populations.
\label{nouv}}
\end{figure}

The multiband data enable us to derive the stellar contents of this system.
Note in Figure~\ref{uvknots} that the UV light along the tail is brightest 
in the farthest half of the (optical) tail. Furthermore, the
brightest UV knot is found at the tip of the tail. It seems that
star formation is more active in the outer regions of tidal tails.
This finding is consistent with the observational evidence 
of TDGs at the tip of the tidal tail in the Antennae galaxy 
\citep[NGC4038/39,][]{hib05} as well as the theoretical predictions
\citep[e.g.,][]{bou06}.

We found three optical objects which appear to be related with the NUV
surface brightness distribution along the tail.  Optical sources
around the tidal tail are indicated as ``TDG1,'' ``TDG2,'' and ``TDG3.''
However, note that none of them is located exactly at the
peaks in the NUV light.  Hence we label the three NUV peaks separately as
``NUV1,'' ``NUV2,'' and ``NUV3.''

We follow two complementary approaches to determine the
color distribution of the sources on the tidal tail. In Section~\ref{sec:pbp},
we present a pixel-by-pixel analysis, which
avoids the ambiguity of matching objects in UV and optical
data. Furthermore, we consider aperture
photometry of the targets, which increases the signal-to-noise
ratio of the photometric data in Section~\ref{sec:ap}.

\subsection{Pixel-by-Pixel Analysis of the Tidal Tail Region}
\label{sec:pbp}

Figure~\ref{colormap} shows the ``FUV $-$ NUV'' and ``NUV $-$ {\it r}'' color maps
derived from pixel magnitude maps in the tidal tail region. Only the
pixels with S/N$_{\rm FUV} > 1$ are shown in ``FUV $-$ NUV'' color map
while the pixels with S/N$_{\rm NUV} > 1$ are presented in ``NUV $-$
{\it r}'' color map.
Surface brightness contours were superimposed on the color maps for
$\mu_{\rm NUV}=26.0, 27.0, 28.0$~mag/arcsec$^2$. The tidal tail shows blue
colors in both cases, suggesting the presence of stellar populations
younger than 1~Gyr (NUV $-$ {\it r} $\sim$ 4 indicates a SSP of
1~Gyr at solar metallicity).

A photometric SED is constructed for each pixel using the pixel magnitude maps.
Stellar population modeling was applied to the pixel SEDs according
to the method described in Section~\ref{sec:model}. The results of
the two-component (with $t_{\rm OC} =12$Gyr) and SSP modeling are presented in
Figures~\ref{pixelmap} and \ref{pixelmap_ssp}. The result for the ``central''
pixel of each source is also presented in Table~\ref{result_tdg} 
(in order to avoid confusion between results using pixel SEDs and 
SEDs based on aperture photometry, we append a ``p'' at the end of their
ID). The ``central'' pixel means a geometric center for TDG 1, 2, 3
and the NUV brightest pixel for NUV 1, 2, 3 actually. 
$1\sigma$ errors are presented in Table~\ref{result_tdg} also.
They indicate that the age and internal extinctions are hardly defined
in SSP modeling.

The results suggest that star formation has taken place along the tidal
tail in the last few hundred Myr. The whole population of ``NUV1p'' 
is young, about 5 Myr old, with a large dust extinction.
All other pixels, except ``TDG3p'', also show significant mass fractions 
of young stars of 7--200 Myr old.
Figure~\ref{pixelgraph} shows the best model fits for ``NUV1p'' and
``TDG1p''. $\chi^2$ contours at the 68 and 90 percent confidence level are
also presented in the right panels showing that parameters are
reasonably well constrained by the data. The open contours of the confidence
levels for the ``NUV1p'' source (top right panel) implies that an upper
limit to the age of the young component (around 5~Myr) can be inferred. 
On the other hand, the two-component modeling of ``TDG3p'' shows a very 
low mass fraction in young stars (0.01 percent), and an old stellar population of a few Gyr
for the SSP modeling. Considering that other parts of the tail are much
younger, it is possible that ``TDG3'' is not associated with NGC\,4922.

\subsection{Aperture Photometry of TDG Candidates}
\label{sec:ap}

\begin{deluxetable}{cr@{}@{}lr@{}@{}lr@{}@{}lr@{}@{}lr@{}@{}lcc}
\tabletypesize{\small}
\tablecaption{Pixel-by-pixel modeling results for the central pixels of our 6 UV/optical sources. 
\label{result_tdg}}
\tablehead{
\colhead{ID} &\multicolumn{2}{c}{$t_{YC,2CM}$} & \multicolumn{2}{c}{$f_{YC}$} &
\multicolumn{2}{c}{$E(B-V)_{2CM}$} &
\multicolumn{2}{c}{$t_{\rm SSP}$} & \multicolumn{2}{c}{$E(B-V)_{\rm SSP}$} & \colhead{RA(J2000)} & \colhead{Dec(J2000)}\\
\colhead{} &\multicolumn{2}{c}{{\scriptsize (Myr)}} & \multicolumn{2}{c}{{\scriptsize (\%)}} & \multicolumn{2}{c}{} & \multicolumn{2}{c}{{\scriptsize (Myr)}} & \multicolumn{2}{c}{} & \colhead{{\scriptsize (hh:mm:ss)}} & \colhead{{\scriptsize (dd:mm:ss)}}
} \startdata
TDG1p & 200 & $^{+110}_{-49.0}$ & 6.5 & $^{+4.3}_{-2.5}$ & 0. & $^{+0.5}$ & 100 & $^{+143}_{-97.4}$ & 0.19 & $^{+0.48}_{-0.064}$ & 13:01:26.62 & 29:17:12.5 \\
TDG2p & 7 & $^{+0.02}_{-0.01}$ & 4.5 & $^{+96}_{-2.9}$ & 0.6 & $^{+0.02}_{-0.2}$ & 7 & $^{+1019}_{-0.01}$ & 0.6 & $^{+0.07}_{-0.3}$ & 13:01:26.56 & 29:17:39.4 \\
TDG3p & 8 & $^{+420}_{-7}$ & 0.01 & $^{+0.2}$ & 0.05 & $^{+0.07}_{-0.05}$ & $2\times10^3$ & $^{+1\times10^4}_{-1\times10^3}$ & 0.2 & $^{+0.3}_{-0.2}$ & 13:01:25.79 & 29:17:07.5\\
NUV1p & 5 & $^{+0.2}_{-4}$ & 100. & $_{-94.4}$ & 0.6 & $^{+0.06}_{-0.5}$ & 5 & $^{+123}_{-4}$ & 0.6 & $^{+0.06}_{-0.5}$ & 13:01:26.47 & 29:17:10.5 \\
NUV2p & 19 & $^{+5.4}_{-4.1}$ & 9. & $^{+91}_{-6}$ & 0.45 & $^{+0.14}_{-0.32}$ & 40 & $^{+101}_{-37}$ & 0.35 & $^{+0.35}_{-0.11}$ & 13:01:26.97 & 29:17:21.5 \\
NUV3p & 200 & $^{+363}_{-110}$ & 2.5 & $^{+5.5}_{-1.5}$ & 0.005 & $^{+0.6}_{-0.005}$ & 6 & $^{+1051}_{-5}$ & 0.6 & $^{+0.1}_{-0.6}$ & 13:01:26.97 & 29:17:36.5 \\
\enddata

\end{deluxetable}

\begin{deluxetable}{cr@{}@{}lr@{}@{}lr@{}@{}lcr@{}@{}lr@{}@{}lc}
\tablecaption{Aperture photometry modeling results for ``TDG1'' and ``NUV1''
\label{result_tdg2}}
\tablehead{
\colhead{ID} & \multicolumn{2}{c}{$t_{YC,2CM}$} & \multicolumn{2}{c}{$f_{YC}$} & \multicolumn{2}{c}{$E(B-V)_{2CM}$} &
\colhead{$M_{2CM}$} & \multicolumn{2}{c}{$t_{\rm SSP}$} &
\multicolumn{2}{c}{$E(B-V)_{\rm SSP}$} &
\colhead{$M_{\rm SSP}$} \\
\colhead{} & \multicolumn{2}{c}{{\scriptsize (Myr)}} & \multicolumn{2}{c}{{\scriptsize (\%)}} & \multicolumn{2}{c}{} &
\colhead{{\scriptsize ($M_{\sun}$)}} & \multicolumn{2}{c}{{\scriptsize (Myr)}} & \multicolumn{2}{c}{} &
\colhead{{\scriptsize ($M_{\sun}$)}}
}
\startdata
TDG1+NUV1 & 4 & $^{+1}_{-3}$ & 20. & $^{+80}_{-17}$ & 0.3 & $^{+0.01}_{-0.1}$ & 4.3$\times10^{6}$ & 3 & $^{+3}_{-2}$ & 0.3 & $^{+0.01}_{-0.1}$ & 9.3$\times10^{5}$ \\
TDG1 & 5 & $^{+0.07}_{-4}$ & 1.2 & $^{+1.8}_{-0.29}$ & 0.16 & $^{+0.024}_{-0.028}$ & 3.1$\times10^{7}$ & 12 & $^{+1.2}_{-11}$ & 0.09 & $^{+0.1}_{-0.08}$ & 1.1$\times10^{6}$ \\
NUV1 & 5 & $^{+0.06}_{-4}$ & 7. & $^{+93}_{-5}$ & 0.09 & $^{+0.03}_{-0.03}$ & 3.4$\times10^{6}$ & 5 & $^{+0.06}_{-4}$ & 0.1 & $^{+0.03}_{-0.03}$ & 3.8$\times10^{5}$ \\
\enddata
\end{deluxetable}

We also carried out aperture photometry of the TDG candidates in the
tidal tail in order to achieve a higher signal-to-noise ratio in their
photometry.
However, ``TDG2'' and ``TDG3'' do not have clear counterparts in the NUV
images, they are just embedded in the extended UV structure. Therefore, 
we cannot apply aperture
photometry for those two objects.
In the case of ``TDG1'', its UV light is blended with ``NUV1'' and 
we can see a faint optical counterpart for ``NUV1'' (the inset image 
in the right panel of Figure~\ref{uvknots}). Both of them show very
blue optical colors ({\it g - r} = 0.12 and -0.26 for ``TDG1'' and ``NUV1,''
respectively). In fact, the pixel-by-pixel analysis in Section~\ref{sec:pbp} 
showed that ``NUV1p'' has significantly younger stellar populations ($\sim 5$ Myr) than 
``TDG1p'' ($\sim 200$ Myr). ``NUV1'' might be significantly contaminating the UV fluxes of 
its surrounding regions. Hence, we can not assume that ``TDG1'' is the sole
optical counterpart of the brightest UV blob at the tip of the tidal
tail. 

Since it was impossible to deblend the UV lights between ``TDG1'' and ``NUV1'' at
the tip of the tail, we integrated the optical SEDs from both objects in
order to match the UV SED. We measured aperture magnitudes of
``TDG1'' and ``NUV1'' with an aperture size, $3 \times$FWHM in the SDSS
data. Also a magnitude of the corresponding UV blob was measured
with an aperture of $3 \times$FWHM from the {\it GALEX} data. We investigated
stellar populations with three kind of SEDs as ``TDG1+NUV1,'' ``TDG1,''
and ``NUV1'' for comparison. 

The results of stellar population modeling are presented
in Table~\ref{result_tdg2}. 
Figure~\ref{tdgsed} shows the best fits and $\chi^2$ contours for three cases. 
The modeling suggests that ``TDG1+NUV1'' has a young stellar population about 
a few Myr old, contributing a substantial mass fraction up to 100 
per cent with considerable internal extinction of 
$E(B-V)=0.3$~mag. If we consider ``TDG1'' and ``NUV1'' separately, 
they show the existence of young stellar population of a few Myr old. 
In all two-component modeling cases the best fits were found when
$t_{\rm OC} = 12$Gyr.
The SSP fits suggest similar results, albeit for larger $\chi^2$ 
than from two-component fits. 

The stellar mass of this system is derived from the $V$-band
luminosity estimated from the $r$-band luminosity using a transformation 
equation extracted from \citet{jor06}. According to
stellar population modeling, the mass of ``TDG1+NUV1'' system was
estimated as 4.3 $\times10^{6} M_{\sun}$ with two-component modeling and
9.2 $\times10^{5} M_{\sun}$ with SSP modeling.  Stellar masses of ``TDG1'' and
``NUV1'' are also calculated separately based on their optical luminosities.
Since the stellar mass-to-light ratio increases with the age of the stellar population,
the mass of a young population derived by two-component fits 
is always greater than the mass estimate from SSP fits. 

The use of the {\it GALEX} UV data played a critical role in the spectral fits.
For example, Figure~\ref{nouv} shows that the best SSP models for ``TDG1+NUV1''
indicates the age of 3~Myr, while the fits without UV data lead to 16~Myr. 
While the $\chi^2$ distributions for these two estimates are far from Gaussian,
they are separated by $3\sigma$ at least. 
Recently, \citet{hib05} and \citet{nef05} have also demonstrated the ability of 
{\it GALEX} data to detect TDGs and to determine the age of young stellar
populations (their estimates were based on SSP modeling using a UV color).

\section{Summary}
\label{sec:summary}
Tidal dwarf galaxies represent one of the
possible progenitors of the general population of dwarf galaxies,
featuring high metallicity and diverse star formation histories. In
this paper we study the tidal tail around a post-merger galaxy,
NGC\,4922, in the Coma cluster. Combining UV and optical photometry,
we discover a few TDG candidates in the tail. In our analysis,
two-component models generally yield a better fit than SSP models for the
tidal tail and TDG candidates, although young populations with a single age
cannot be ruled out. Also TDG candidates showed a better fit with models
with higher metallicity for a young component.
According to the pixel-by-pixel and the aperture
analyses, two optical TDG candidates and three NUV peaks show
predominantly young stellar populations of a few Myr of age, suggesting
that recent star formation occurred almost simultaneously along the
tidal tail. The result also corresponds with the recent work on
the interacting galaxy Arp\,305 in which the age of TDG candidates was
$\sim$6 Myr \citep{han09}.

The pixel-by-pixel analysis adopted in this study is relatively new to
the field. 
With the new deep and multiband photometric data, one can
now perform statistically meaningful analyses pixel by pixel.  A
clear advantage of the pixel-by-pixel analysis is that it allows us to
make spatially resolved age maps, instead of estimates for an integrated region.
Hence we can investigate the star formation history of 
the galaxy in connection with the merging event. The Coma cluster is a bit too far
for such a detailed investigation. Closer cluster environments,
such as Virgo and Fornax, provide better opportunities. 

The UV data proved once again to be powerful for searching for young
starbursts.  The {\it GALEX} UV data on hundreds of nearby galaxies, mainly
but not exclusively in the NGS survey mode, will provide other
interesting opportunities without doubt.

{\it Facilities:} \facility{$\it{GALEX}$}, \facility{SDSS}, \facility{Perkins (PRISM)}

\acknowledgements
We are indebted to Giuseppe Gavazzi and Alessandro Boselli for supporting
our use of the {\it GALEX} data made public after their guest investigation acquired
them but prior to their use.
We thank the anonymous referee for various clarifications.
This research was supported by Basic Science Research Program through
the National Research Foundation of Korea (NRF) funded by the Ministry of
Education, Science and Technology (Doyak 20090078756).
S.K.Y. also acknowledges support from Korea Astronomy and Space Science Institute.
We have used the {\it GALEX} UV data obtained from the Multimission Archive
at the Space Telescope Science Institute (MAST).
{\it GALEX} is operated for NASA by the California Institute of Technology
under NASA contract NAS5-98034. 
We are grateful to the Lowell Observatory for granting observing time and
hospitality during our visit.

\end{document}